\documentclass[prl,twocolumn,showpacs,superscriptaddress]{revtex4-1}
\usepackage{amsmath,amssymb}
\usepackage{graphicx,color}
\usepackage{times}
\newcommand{\bra}[1]{ \langle #1 |}
\newcommand{\ket}[1]{| #1 \rangle}
\newcommand{\bk}[2]{\left \langle #1 | #2 \right \rangle}
\newcommand{\aver}[1]{\left \langle #1 \right \rangle}
\renewcommand{\Im}{\textrm{Im}}

\newcommand{\Heff}{\mathcal{H}_{\mathrm{eff}}}

\begin{document}

\title{Statistics of resonance width shifts as a signature of eigenfunction nonorthogonality}
\author{Yan V. Fyodorov}
 \affiliation{Queen Mary University of London, School of Mathematical Sciences, London E1 4NS, United Kingdom}
\author{Dmitry V. Savin}
 \affiliation{Department of Mathematical Sciences, Brunel University, Uxbridge, UB8 3PH, United Kingdom}
\published{3 May 2012 in \texttt{Phys.\,Rev.\,Lett.\,\textbf{108},\,184101\,(2012)}}

\begin{abstract}
We consider an open (scattering) quantum system under the action of a perturbation of its closed counterpart. It is demonstrated that the resulting shift of resonance widths is a sensitive indicator of the nonorthogonality of resonance wavefunctions, being zero only if those were orthogonal. Focusing further on chaotic systems, we employ random matrix theory to introduce a new type of parametric statistics in open systems, and derive the distribution of the resonance width shifts in the regime of weak coupling to the continuum.
\end{abstract}
\maketitle

The classical question of how energy levels of a quantum system get shifted under the action of a perturbation kept attracting renewed attention during the last two decades, mostly due to the established universality of such a parametric motion for systems with chaotic dynamics or intrinsic disorder \cite{simo93b,smol03a}. In particular, the distributions and correlation functions of  parametric derivatives of energy levels (``level velocities'') \cite{simo93b,smol03a,krav92,*fyod94} and their second derivatives (``level curvatures'') \cite{zakr93a,*v.oppe94,*fyod95} were found explicitly using the methods of random matrix theory (RMT) \cite{guhr98}, and also verified, e.g., in microwave billiard experiments \cite{bart99}. The other reason for such an interest is the recent development of the fidelity concept as the measure of the susceptibility of internal dynamics to perturbations \cite{gori06r,*jacq09}.

Experimentally, the energy levels are mostly accessible by means of a scattering setup \cite{Stoeckmann}. From such a viewpoint, parametric dependencies of scattering characteristics, like phase shifts and time delays \cite{fyod96a,*fyod97a}, conductances \cite{alha96b,*brou97b} and $S$ matrix elements \cite{mac94e,*hoeh08,*koeb10} were under intensive study. As to the parental discrete energy levels, they are converted into the resonances with finite lifetimes, since the original closed system becomes open (unstable). Such resonances manifest themselves in the energy-dependent $S$ matrix as its poles in the complex energy plane, and can be analytically described as the complex eigenvalues of an effective non-Hermitian Hamiltonian \cite{verb85,soko89,fyod97}. Notably, the corresponding eigenfunctions are not orthogonal in the conventional sense but rather form a biorthogonal system. Their nonorthogonality plays an important role in many applications, e.g., describing interference in neutral kaon systems \cite{bell66}, influencing branching ratios of nuclear cross sections \cite{soko97i}, and yielding excess noise in open laser resonators \cite{pete79,schom00a}. It also features in decay laws of quantum chaotic systems \cite{savi97} and in dissipative quantum chaotic maps \cite{schom04,*keat06}.

In such a context the question of parametric motion of resonances and associated resonance states in open systems arises very naturally, but to the best of our knowledge has never been properly addressed. Our goal here is to begin filling in that gap by considering universal statistics of the shifts in the resonance widths under a generic perturbation in chaotic systems. In particular, we will demonstrate that such shifts are a clear manifestation of eigenstate nonorthogonality, thus providing a promising way to probe this spatial characteristics by purely spectroscopic tools. To this end let us stress that statistics of complex poles and lifetimes in chaotic systems can be verified via accurate scattering experiments in microwave billiards \cite{kuhl08} or photonic crystals \cite{difa11}. It can be also extracted from realistic computer simulations of quantum graphs \cite{kott00}, semiconductor superlattices \cite{glue02}, dielectric microresonators \cite{schom09} or system of randomly interacting fermions \cite{cela11}. As to the theoretical framework, it mainly relies on studying the relevant non-Hermitian RMT \cite{soko89,fyod97}, understanding of which has substantially improved over the last two decades; see, e.g., Ref.~\cite{fyod03r} and references therein. Note that the spatial properties related to the associated bi-orthogonal eigenvectors are known to a much lesser extent \cite{schom00a,fyod02,*brou03,savi06b,poli09b}.

As is well known \cite{verb85,soko89,fyod97} (see also \cite{mitc10,*fyod11ox} for most recent reviews), resonance phenomena involving a group of $N$ interfering resonances can be adequately described in terms of the  effective non-Hermitian Hamiltonian
\begin{equation}\label{Heff}
 \Heff = H - i \gamma WW^\dag\,,
\end{equation}
which governs the dynamics of the open system. Here, the Hermitian $N{\times}N$ matrix $H$ corresponds to the Hamiltonian of the closed counterpart, whereas the entries  $W_n^c$ of the rectangular $N{\times}M$ matrix $W$ are the decay amplitudes that describe coupling of $N$ discrete energy levels, labeled by $n$, to $M$ decay channels, labeled by $c$. The coupling strength to the continuum is controlled by the dimensionless positive constant $\gamma$, with $\gamma\ll 1$ ($\gamma=1$) being the particular case of weak (perfect) coupling. The eigenvalue problem for the full non-Hermitian matrix $\Heff$ reads as follows
\begin{equation}\label{eigen0}
 \Heff\ket{R_n} = \mathcal{E}_n\ket{R_n}\quad\mathrm{and}\quad
 \bra{L_n}\Heff = \mathcal{E}_n\bra{L_n}
\end{equation}
and determines its complex spectrum $\mathcal{E}_n=E_n-\frac{i}{2}\Gamma_n$. Here $E_n$ stands for the resonance positions (energies) and $\Gamma_n>0$ denotes the corresponding widths \cite{note1}. The set of the associated left and right eigenvectors (resonance wavefunctions) satisfies the conditions of biorthogonality, $\bk{L_n}{R_m}=\delta_{nm}$, and completeness, $\sum_{n=1}^N\ket{R_n}\bra{L_n}=1$.

In such a framework a perturbation of the closed counterpart can be modeled by the term $\alpha V$, where $V$ is a Hermitian $N\times N$ matrix and the real  constant $\alpha$ is to control the perturbation strength. The resonance parameters for the perturbed open system are then to be determined from solving the following spectral problem for the right eigenstates,
\begin{equation}\label{eigen}
 (\Heff +\alpha V)\ket{R_n'} = \mathcal{E}_n'\ket{R'_n}
\end{equation}
and a similar problem for the left eigenstates $\bra{L'_n}$. In the case of the weak perturbation, $\alpha\ll1$, one can follow the standard perturbation theory routine and seek each of the two eigenvectors corresponding to the new eigenvalue $\mathcal{E}_n'$ as an expansion over non-perturbed basis of $\Heff$, with necessary modifications induced by bi-orthogonality \cite{note2}. To the first order in $\alpha$ this readily yields the expression for the shift of the $n$th resonance in the form
\begin{equation}\label{shift}
 \delta\mathcal{E}_n \equiv \mathcal{E}_n'-\mathcal{E}_n = \alpha\bra{L_n}V\ket{R_n},
\end{equation}
generalizing the standard result to the non-Hermitian case.

The resonance shift (\ref{shift}) contains both real and imaginary parts, since $\bra{L_n}\neq(\ket{R_n})^\dag$ in general \cite{note3}. At this point we stress that a nonzero value of the imaginary part of $\delta\mathcal{E}_n$ is induced solely due to  nonorthogonality of the resonance states. This fact becomes apparent in the following equivalent representation for the resonance width shift $\delta\Gamma_n\equiv-2\Im(\delta\mathcal{E}_n)$:
\begin{eqnarray}\label{dgam}
 \delta\Gamma_n &=& i\alpha(\bra{L_n}V\ket{R_n}-\bra{R_n}V\ket{L_n}) \nonumber
 \\ &=& i\alpha\sum_{m}(U_{nm}V_{mn}-V_{nm}U_{mn})\,,
\end{eqnarray}
where we have made use of the completeness condition to expand $\ket{L_n}=\sum_m\ket{R_m}\bk{L_m}{L_n}$ and also introduced $V_{nm}=\bra{R_n}V\ket{R_m}$ and $U_{nm}=\bk{L_n}{L_m}$. Since by construction $V_{nm}=V_{mn}^{*}$ and $U_{nm}=U_{mn}^{*}$, only the terms with $m\neq n$ contribute to the sum (\ref{dgam}). The matrix $U$ is just the Bell-Steinberger nonorthogonality matrix \cite{bell66} (see also a compact description in \cite{soko89}), and $U_{nm}\neq\delta_{nm}$ in general. Thus, the resonance width shift (\ref{dgam}) would generically vanish only if the resonance states were orthogonal \cite{note4}, being thus a sensitive indicator of their nonorthogonality.

In the rest of the Letter, we concentrate on the regime of weak coupling to the continuum, $\gamma\ll1$, which permits complete analytical investigation, and is the one that is the most easily realized experimentally. Under such an assumption  the non-Hermitian part $i \gamma WW^\dag$ of $\Heff$
can be treated as the perturbation of the Hermitian part $H$. To the leading order in $\gamma$ the resonance positions coincide with the energy levels of the closed system, $H\ket{n}=E_n\ket{n}$, whereas the resonance widths are given by expression
$\Gamma_n = 2\gamma\bra{n}WW^\dag\ket{n}=2\gamma\sum_{c=1}^M|W_n^c|^2$.
Similarly, the right eigenvectors of $\Heff$ are readily found to be $\ket{R_n}=\ket{n}-i\gamma\sum_{m\neq n}\frac{(WW^\dag)_{mn}}{E_n-E_m}\ket{m}$,
while the left ones read $\bra{L_n}=\bra{n}-i\gamma\sum_{m\neq n}\frac{(WW^\dag)_{nm}}{E_n-E_m}\bra{m}$. Substituting such expressions into Eq.~(\ref{shift}), one finds that to the leading order in both $\alpha$ and $\gamma$ the energy shift is given by the standard expression for the closed system $\delta E_n = \alpha\bra{n}V\ket{n}$, whereas the shift in the resonance width is determined by
\begin{equation}\label{dgam1}
  \delta\Gamma_n = 2\alpha\gamma\sum_{m\neq n}\frac{\bra{m}G_n\ket{m}}{E_n-E_m}\,,
\end{equation}
where $G_n$ stands for the following Hermitian operator:
\begin{equation}\label{G_n}
  G_n=WW^\dag\ket{n}\bra{n}V+V\ket{n}\bra{n}WW^\dag\,.
\end{equation}

Aiming to  describe the universal statistics of the resonance shifts for generic chaotic systems (e.g., open billiards, quantum dots, quantum graphs), we follow the standard paradigm \cite{guhr98,Stoeckmann} and model $H$ by a random $N\times N$ matrix drawn from the Gaussian orthogonal (GOE) or unitary (GUE) ensemble, depending on the presence or absence of time-reversal symmetry, respectively. Universal fluctuations are then expected to occur in the limit $N\gg 1$ at the local scale of the order of the mean level spacing $\Delta\sim1/N$. Without loss of generality, one can restrict the consideration to the spectrum center, where $\Delta=\lambda\pi/N$ and $2\lambda$ is the semicircle radius ($\Delta$ needs to be rescaled if $E\neq0$). As to the coupling amplitudes, they can be taken \cite{soko89} as real (GOE, $\beta{=}1$) or complex (GUE, $\beta{=}2$) Gaussian random variables with zero mean and the variance $\aver{W_n^aW_m^{b*}}=(\Delta/\pi)\delta_{nm}\delta^{ab}$, the final results being model-independent provided the number of channels $M\ll N$. This readily yields the well-known result for the resonance width distribution in terms of $\chi^2_\nu$  distribution with $\nu=M\beta$ degrees of freedom (Porter-Thomas expression at $M=1$ and $\beta=1$),
\begin{equation}\label{PTdis}
 P_M^{(\beta)}(\kappa) = \frac{(2/\beta)^{M\beta/2} }{ \Gamma(M\beta/2)}\kappa^{M\beta/2-1}e^{-\beta\kappa/2}\,,
\end{equation}
where $\kappa_n=\frac{\pi\Gamma_n}{2\gamma\Delta}$ stands for the width measured in units of the mean partial width (per channel). Distribution (\ref{PTdis}) has the mean value $\aver{\kappa}=M$ and variance $\mathrm{var}(\kappa)=\frac{2}{M\beta}\aver{\kappa}^2$.

In the limit $N\gg1$, the rescaled matrix elements $v^{(n)}_m=N\bra{m}V\ket{n}/\sqrt{\mathrm{Tr}(V^2)}$ of the perturbation in the eigenbasis of $H$ become normally distributed random variables \cite{fyod95b}. This results in the Gaussian distribution of the energy shifts $\delta E_n$ (i.e. ``level velocities'') with $\mathrm{var}(\delta E_n)=\frac{2\alpha^2}{\beta N^2}\mathrm{Tr}(V^2)$. At the same time the width shifts $\delta\Gamma_n$  must clearly have much less obvious distribution due to the nontrivial structure of Eq.~(\ref{dgam1}). To find the latter distribution explicitly, we first scale the width shifts in the natural units to get rid of the model-dependent features, and introduce
 \begin{equation}\label{y_n}
 y_n = \frac{\delta\Gamma_n}{2\gamma\sqrt{2\beta\,\mathrm{var}(\delta E_n)}}\,.
\end{equation}
In close analogy with the case of the closed systems such a rescaling is expected to capture universal (local) fluctuations related to the parametric motion in generic open systems. At the next step it is instructive to follow Ref.~\cite{poli09b} and treat the scalar product $\frac{\pi}{\Delta\sqrt{\kappa_n}}(WW^\dag)_{mn}=z_m^{(n)}$ as a projection of the $M$-dimensional vector of the decay amplitudes $\sqrt{\frac{\pi}{\Delta}}W_m$ onto the direction determined by the vector $W_n$ at given $n\neq m$. As a result, the expression for the rescaled width shifts takes the following convenient form:
\begin{equation}
 y_n  = \frac{\sqrt{\kappa_n}}{\pi} \sum_{m\neq n}\frac{\Delta\mathrm{Re}(z^{(n)*}_m v^{(n)}_m)}{E_n-E_m}\,.
\end{equation}
The advantage of such a parametrization is that the corresponding modula and angles turn out to be statistically independent \cite{soko89}. The projections $z^{(n)}_m$ can then be shown to be normally distributed random variables (real at $\beta=1$ and complex at $\beta=2$) at any $M\geq 1$ \cite{poli09b}.

We now compute the probability distribution ${\cal P}_M(y)$ of the rescaled width shifts (at the spectrum centre) defined as
\begin{equation}\label{P_y}
 {\cal P}_M(y) = \Delta\aver{\sum_{n=1}^N\delta(E_n)\overline{\delta(y-y_n)}}\,,
\end{equation}
where the angular brackets denote the spectral average over both energies $E_n$ and widths $\kappa_n$, whereas the bar stands for the averaging over the quantities $z^{(n)}_m$ and $v^{(n)}_m$, all of them being statistically independent. The latter task can be easily performed by considering the Fourier transform, yielding
\begin{equation}
 \overline{e^{-i\omega y_n}} = \prod_{m\neq n}\frac{|E_n-E_m|^{\beta}}{[(E_n-E_m)^2+
 \kappa_n(\omega\Delta/\pi\sqrt{\beta})^2]^{\beta/2}}\,.
\end{equation}
Substituting this expression back to Eq.~(\ref{P_y}), one can then make use of the known explicit form of the joint probability function of all eigenvalues to integrate out the $n$th eigenvalue. The distribution can be finally represented as follows
\begin{equation}\label{shiftdist}
{\cal P}_M(y) =
   \int_0^{\infty} \frac{d\kappa}{\sqrt{\kappa}}\, P_M^{(\beta)}(\kappa)\, \phi^{(\beta)}\left( \frac{y}{\sqrt{\kappa}}\right)\,,
\end{equation}
where the width distribution $P_M^{(\beta)}(\kappa)$ is given by Eq.~(\ref{PTdis}) and the function
$\phi^{(\beta)}(y) = \int_{-\infty}^{\infty}\frac{d\omega}{2\pi}e^{i\omega y} C_{N-1}^{(\beta)}(\omega)$
is the Fourier-transform of the following ratio of spectral determinants for the (GOE or GUE) matrix $H_1$ of the reduced size $N-1$:
\begin{equation}\label{specdet}
 C_{N-1}^{(\beta)}(\omega) = c_{N-1}^{(\beta)}
 \aver{ \frac{\det(H_1)^{2\beta}}{\det[H_1^2+(\omega\Delta/\pi\sqrt{\beta})^2]^{\beta/2}} }\,.
\end{equation}
The constant $c_{N-1}^{(\beta)}=\aver{\det|H_1|^{-\beta}}$ ensures $C_{N-1}^{(\beta)}(0)=1$ that automatically guarantees the normalization of distribution (\ref{shiftdist}) to unity.  Objects similar to Eq.~(\ref{specdet}) naturally  arise in the analysis of weakly open chaotic systems as a result of separating independent fluctuations in spectra and in wavefunctions.  Following the methods developed in \cite{schom00a} for $\beta=1$ and in \cite{stra03} for $\beta=2$, we have been able to calculate the limiting expressions at $N\gg1$, with the explicit result being
\begin{equation}\label{specdet12}
  \begin{array}{ll}
    C^{(1)}_{\infty}(\omega) = \frac{1}{3} \left[|\omega|K_1(|\omega|)+\omega^2K_2(|\omega|)\right],
    \\[1ex]
    C^{(2)}_{\infty}(\omega) = e^{-2|\omega|} \left(1+2|\omega|+\frac{4\omega^2}{3}+\frac{|\omega|^3}{3}\right),
  \end{array}
\end{equation}
where $K_{\nu}(x)$ stands for the modified Bessel (Macdonald) function. Taking the Fourier transform, we finally arrive at
\begin{equation}\label{fourier}
 \phi^{(\beta)}\left(y\right) = \left\{
 \begin{array}{ll}
  \displaystyle \frac{4+y^2}{6(1+y^2)^{5/2}}\,, & \beta{=}1 \\[2ex]
  \displaystyle \frac{35+14y^2+3y^4}{12\pi(1+y^2)^4}\,, & \beta{=}2
 \end{array}
 \right. .
\end{equation}

\begin{figure}[t]
 \includegraphics[width=0.45\textwidth]{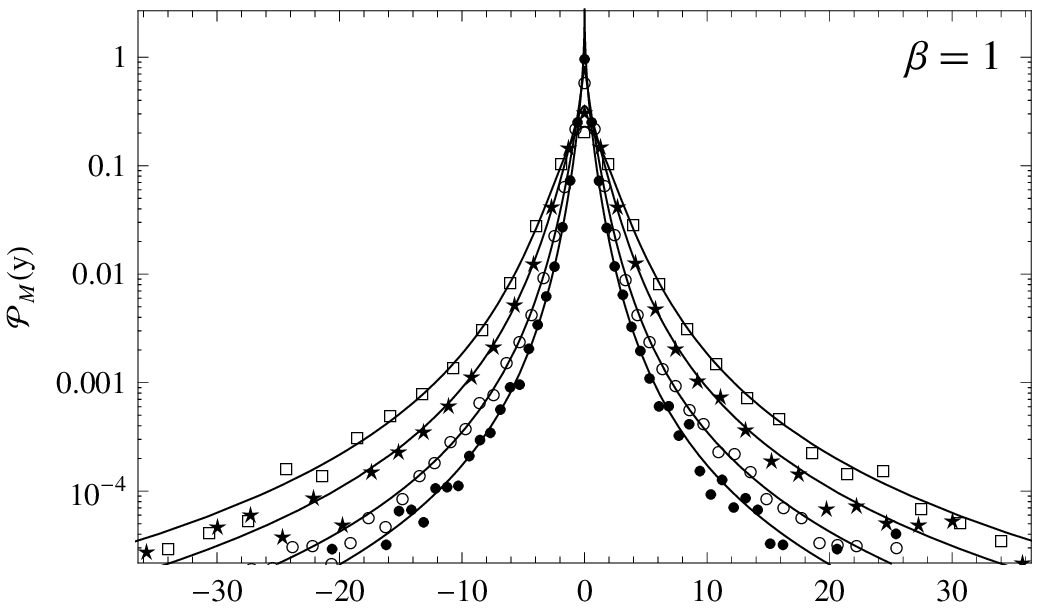}\\
 \includegraphics[width=0.45\textwidth]{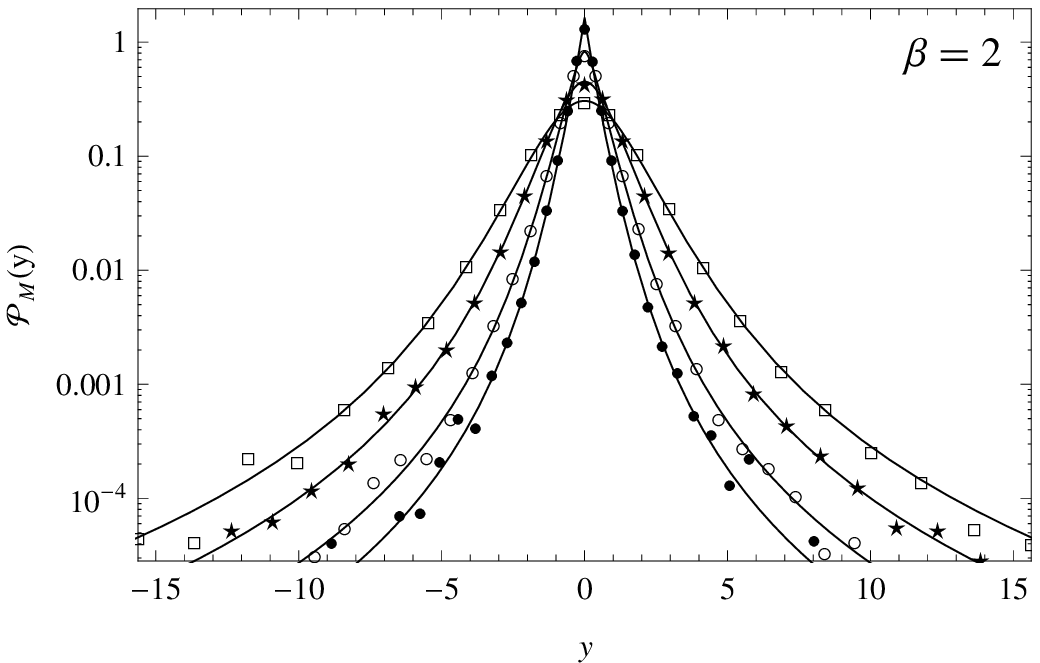}\\
 \caption{Distributions of the resonance width shifts for weakly open chaotic systems with preserved ($\beta{=}1$, top) or broken ($\beta{=}2$, bottom) time-reversal symmetry at $M=1\,(\bullet), 2\,(\circ), 5\,(\star)$ and $10\,(\Box)$ open channels. The solid lines show our analytical prediction, Eqs.~(\ref{shiftdist}) and (\ref{fourier}). The symbols stand for numerics with $2000$ realizations of $250{\times}250$ GOE ($\beta{=}1$) or GUE ($\beta{=}2$) random matrices (only 25 levels around the spectrum centre were taken for each realization).}
\end{figure}
Combination of Eq.~(\ref{shiftdist}) with Eqs.~(\ref{PTdis}) and (\ref{fourier}) completely solves the problem of universal statistics of the resonance width shifts in a chaotic system weakly coupled to the continuum via $M$ equivalent channels. We see that far tails of the distribution decay as $\mathcal{P}_M(y)\propto y^{-(\beta+2)}$ due to the Wigner-Dyson level repulsion at small energy level separations, the feature which such a distribution shares with that for level curvatures \cite{zakr93a}.  In contrast to the level curvature distribution, the broadness of the width shift distribution (\ref{shiftdist}) can be additionally controlled and is proportional to $\sqrt{M}\sim\sqrt{\mathrm{var}(\kappa)}$. Physically, this gives the variance of widths the role of a universal parameter that controls the degree of nonorthogonality in weakly open chaotic systems \cite{savi06b,poli09b}. In the limit of many weakly open channels $M\gg1$ (but still $M\ll N$) the widths cease to fluctuate, so distribution (\ref{PTdis}) becomes very narrow and peaked around its mean value $\aver{\kappa}=M$.  In such a limit the width shifts are still widely distributed, with the probability density for the scaled variable $\tilde{y}=y/\sqrt{M}$ being given just by the function $\phi^{(\beta)}(\tilde{y})$, Eq.~(\ref{fourier}). These findings are illustrated on Fig.~1 which shows the distribution $\mathcal{P}_M(y)$ at several values of $M$. Also shown are the results of straightforward numerical simulations of the width shifts (\ref{y_n}) with GOE/GUE random matrices, the agreement being flawless.

In summary, we have shown that the change in the resonance widths due to external perturbation is a sensitive indicator of eigenfunction nonorthogonality in open quantum or wave systems, and have computed its distribution analytically for weakly open chaotic systems with preserved or broken time-reversal symmetry. Our predictions can be verified, e.g., in experiments with reverberant dissipative bodies \cite{lobk00a}, microwave cavities \cite{bart05a,*poli10}, and elastic plates \cite{xeri09}, where other aspects of nonorthogonality were investigated. Potentially, they may also have implications for the other type of non-Hermitian systems, those with $PT$-symmetry, which is a rapidly developing field \cite{bend07}. We also note a link to a more formal concept of pseudospectra of non-selfadjoint operators considered mostly in mathematical literature \cite{tref97}. The generalization of our results to the case of arbitrary modal overlap is another challenging problem to consider in future studies.

YVF acknowledges financial support from the EPSRC Grant No. EP/J002763/1 ``Insights into Disordered Landscapes via Random Matrix Theory and Statistical Mechanics''.


\begin{thebibliography}{48}
\bibitem{note1}
    The positivity of the widths is automatically ensured due to the factorized structure of the anti-Hermitian part of $\Heff$ which, in its turn, is a direct consequence of the unitarity constraint \cite{soko89}.
\bibitem{note2}
    Substituting $\ket{R_n'}=\sum_{m=1}^N r^{(n)}_m\ket{R_m}$ into Eq.~(\ref{eigen}) and taking the  scalar product with $\bra{L_k}$, one arrives at the system of linear equations, $\sum_{m}r^{(n)}_m[(\mathcal{E}_m-\mathcal{E}_n')\delta_{mk}+\alpha\bra{L_k}V\ket{R_m}]=0$, which can be solved iteratively in powers of $\alpha$. At $k=n$ it gives the first order result (\ref{shift}), whereas $k\neq n$ yields the coefficients $r^{(n)}_{k\neq n} = \alpha\bra{L_k}V\ket{R_n}/(\mathcal{E}_n-\mathcal{E}_k)$. Likewise, for the perturbed left eigenvector $\bra{L_n'}=\sum_{m=1}^N l^{(n)}_m\bra{L_m}$ one finds $l^{(n)}_{k\neq n} = \alpha\bra{L_n}V\ket{R_k}/(\mathcal{E}_n-\mathcal{E}_k)$, and $r^{(n)}_{n}=l^{(n)}_{n}=1$.
\bibitem{note3}
    Note that the cumulative shift is still purely real (for $V^\dag=V$), $\sum_n\delta\mathcal{E}_n = \alpha\sum_n\mathrm{Tr}(V\ket{R_n}\bra{L_n}) = \alpha \mathrm{Tr}(V)$. As a result, the shifts of the resonance widths must satisfy the sum rule $\sum_n\delta\Gamma_n = 0$. Actually, the above result holds at arbitrary $\alpha$, due to $\sum_n\mathcal{E}_n'=\mathrm{Tr}(\Heff+\alpha V)=\sum_n\mathcal{E}_n+\alpha\mathrm{Tr}(V)$.
\bibitem{note4}
    Another option would be a special symmetry that satisfies commutation relations in (\ref{dgam}), a rather peculiar case that we do not consider here.
\end{thebibliography}

%

\end{document}